\pdfoutput=1
\documentclass[a4paper,12pt]{article}
\usepackage[a4paper, top=1.6cm, bottom=1.6cm, left=1.6cm, right=1.6cm]%
{geometry}
\usepackage{natbib} 
\usepackage[utf8]{inputenc}
\usepackage{setspace}
\usepackage{amsmath,epsfig,epsf,psfrag, url, setspace, graphicx,natbib,mathtools,subfigure,comment}
\usepackage{a4wide,amsmath, amsthm, amssymb, multirow}
\usepackage{xcolor,colortbl}
\usepackage{amsfonts}
\usepackage{verbatim}
\usepackage{arydshln} 
\usepackage{placeins} 
\providecommand{\keywords}[1]
{
  \small	
  \textbf{\textit{Keywords---}} #1
}
\usepackage{bm}
\usepackage{makecell}
\usepackage[flushleft]{threeparttable} 
\usepackage{bm}
\usepackage{graphicx}
\usepackage{placeins}
\usepackage{amsmath}
\usepackage{amsfonts} 
\usepackage{commath} 
\usepackage{stackengine}
\usepackage[T1]{fontenc} 
\stackMath
\newcommand\tsup[2][2]{
 \def\useanchorwidth{T}%
  \ifnum#1>1%
    \stackon[-.5pt]{\tsup[\numexpr#1-1\relax]{#2}}{\scriptscriptstyle\sim}%
  \else%
    \stackon[.5pt]{#2}{\scriptscriptstyle\sim}%
  \fi%
}
\usepackage{tabularx, booktabs}
\usepackage{multirow}
\usepackage[flushleft]{threeparttable}
\usepackage{makecell}
\usepackage{adjustbox}
\usepackage{appendix}
\usepackage[figuresright]{rotating} 
\usepackage{hyperref}
\usepackage{cleveref}
\usepackage{times} 

\title{Bayesian Spatial Models for Voxel-wise Prostate Cancer Classification Using Multi-parametric MRI Data}
\author{Jin Jin$^{1,\ast}$, Lin Zhang$^2$, Ethan Leng$^3$,\\ Gregory J. Metzger$^4$, and Joseph S. Koopmeiners$^{2}$\\
$^{\ast}$jjin31@jhu.edu\\
$^1$Department of Biostatistics, Johns Hopkins Bloomberg\\School of Public Health, Baltimore, MD, USA\\
$^2$Division of Biostatistics, School of Public Health, University of Minnesota,\\Minneapolis, MN, USA\\
$^3$Department of Biomedical Engineering, University of Minnesota,\\Minneapolis, MN, USA\\
$^4$Department of Radiology, Center for Magnetic Resonance Research, \\University of Minnesota, Minneapolis, MN, USA}
\date{}

\begin{document}

\maketitle

\begin{abstract}
Multi-parametric magnetic resonance imaging (mpMRI) plays an increasingly important role in the diagnosis of prostate cancer.
Various computer-aided detection algorithms have been proposed for automated prostate cancer detection by combining information from various mpMRI data components. 
However, there exist other features of mpMRI, including the spatial correlation between voxels and between-patient heterogeneity in the mpMRI parameters, that have not been fully explored in the literature but could potentially improve cancer detection if leveraged appropriately. 
This paper proposes novel voxel-wise Bayesian classifiers for prostate cancer that account for the spatial correlation and between-patient heterogeneity in mpMRI. 
Modeling the spatial correlation is challenging due to the extreme high dimensionality of the data, and we consider three computationally efficient approaches using Nearest Neighbor Gaussian Process (NNGP), knot-based reduced-rank approximation, and a conditional autoregressive (CAR) model, respectively.
The between-patient heterogeneity is accounted for by adding a subject-specific random intercept on the mpMRI parameter model.
Simulation results show that properly modeling the spatial correlation and between-patient heterogeneity improves classification accuracy. 
Application to in vivo data illustrates that classification is improved by spatial modeling using NNGP and reduced-rank approximation but not the CAR model, while modeling the between-patient heterogeneity does not further improve our classifier. Among our proposed models, the NNGP-based model is recommended considering its robust classification accuracy and high computational efficiency.\\
\end{abstract} \hspace{10pt}



\section{Introduction}\label{intro}
Multi-parametric magnetic resonance imaging (mpMRI) 
has continued to play an important role in the detection of prostate cancer (Kurhanewicz et al., 2008; Dickinson et al., 2011). 
Conventional manual diagnosis using mpMRI 
does not realize its full potential due to demonstrated radiologist and experience-dependent variability 
(Rosenkrantz et al., 2013; Garcia-Reyes et al., 2015).
To avoid such shortcomings, various computer-aided detection (CAD) algorithms have been proposed, seeking automated cancer detection using different methods to combine information from various MRI parameters, such as linear and nonlinear regressions, clustering methods, support vector machine, ensemble learning approaches and naive Bayes 
(Niaf et al., 2012; Shah et al., 2012; Vos et al., 2012; Peng et al., 2013; Khalvati, Wong and Haider, 2015). 
However, these existing methods have been found insufficient, either lacking the flexibility to model important features of the mpMRI data, or are overly flexible, resulting in poor performance with the sample sizes found in prostate cancer imaging studies. 
Previously, it has been observed that there exists substantial difference in both the distribution of the observed mpMRI parameters and cancer prevalence between the two regions of the prostate: the peripheral zone (PZ) and central gland (CG). 
Recently, Jin et al. (2018) proposed to model the regional heterogeneity using a Bayesian hierarchical modeling framework, which has a flexible structure allowing complex distributional assumptions for both the predictors (mpMRI parameters) and the outcome (cancer status) that cannot be easily accounted for by the existing CAD algorithms. 

In addition to the regional variability, 
mpMRI data also show substantial spatial correlation in the voxel-wise cancer status, as well as between-patient heterogeneity in the mpMRI parameters which is possibly due to registration error or variability in patients' physical conditions.
Although Jin et al. (2018) proposed to account for spatial correlation between voxels using post hoc spatial smoothing, 
systematic modeling of these various mpMRI features has not been formally investigated in the context of fully automated, voxel-wise prostate cancer detection. 
The major challenge associated with modeling the spatial correlation is the extreme high dimensionality of the data. 
Typically, 
each mpMRI image has thousands of voxels, which must be modeled simultaneously over multiple images for model development. Spatial modeling requires inverting large spatial covariance matrices that typically requires $\sim n^3$ floating point operations and storage of order $n^2$, with $n$ denoting the number of voxels in an image, which is computationally infeasible for our motivating data set.

There are two general approaches for modeling 
large spatial data sets.
One is sparse approximation, which introduces sparsity in the spatial covariance or precision matrix. Methods include 
covariance tapering assuming compactly supported covariance functions (e.g. 
Kaufman, Schervish and Nychka, 2008), and approximating the likelihood by the product of lower dimensional conditional densities (e.g. Vecchia, 1988; Stein, Chi and Welty, 2004), Markov random fields (e.g. Rue and Held, 2005), or composite likelihoods (e.g. Eidsvik et al., 2014). 
More recently, Datta et al. (2016) proposed a Nearest Neighbor Gaussian Process (NNGP) for fully process-based modeling of large spatial data sets, which approximates the likelihood of a spatial process by the product of conditional densities between nearest neighbors. 
The other general 
approach is reduced-rank approximation, which 
constructs spatial processes on a lower-dimensional subspace.
Methods include predictive process models (Banerjee et al., 2008; Finley, Banerjee and McRoberts, 2009), low rank splines or basis functions (Cressie and Johannesson, 2008) and kernel convolutions (Higdon, 2002). 

This paper aims to improve voxel-wise classification of prostate cancer by systematically modeling the between-patient variability in the mpMRI parameters and spatial correlation in the voxel-wise cancer status via Bayesian hierarchical modeling,
which can potentially better localize and determine the extent of cancer lesions, providing enhanced guidance for needle biopsy and focal therapy in clinical practice.
Our proposed models build on the work of Jin et al. (2018), which accounts for regional variability in the various data components, and will serve as the ``baseline'' to evaluate the performance of our proposed models. The between-patient heterogeneity is incorporated by specifying a subject specific random intercept for the mpMRI parameters. 
Unlike the post hoc spatial smoothing technique used by Jin et al. (2018), we propose to formally model the spatial correlation structure in the voxel-wise cancer status via three scalable spatial modeling approaches: a sparse approximation approach using the NNGP proposed by Datta et al. (2016), a reduced-rank approximation approach using the knot-based method in Banerjee et al. (2008), and Gaussian Markov random fields implemented via the conditional autoregressive (CAR) model. 
Our simulation results illustrate that the proposed models substantially improve classification accuracy both by incorporating the between-patient heterogeneity and by modeling the spatial correlation. 
Application to our motivating data set demonstrates improvement due to modeling the spatial correlation using the NNGP and reduced-rank approximation but not the CAR model, while modeling the between-patient heterogeneity does not further improve classification accuracy.
Among our proposed scalable spatial modeling approaches, the NNGP-based approach is recommended considering its robust classification accuracy with respect to spatial correlation pattern and high computational efficiency.

The remainder of the paper proceeds as follows. 
In Section \ref{data}, we describe our motivating data set. In Section \ref{method}, we discuss methods development and implementation. 
In Section \ref{simulation}, we present simulation results illustrating the performance of the various models discussed in Section \ref{method}, and in Section \ref{application}, we assess model performance on our motivating data set.
We conclude with a discussion of the model properties and potential extensions in Section \ref{discussion}.

\section{Overview of the mpMRI data and notations}\label{data}
\begin{figure}[ht!]
\begin{center}
	\centerline{
    \includegraphics[width=160pt,height=10.8pc]{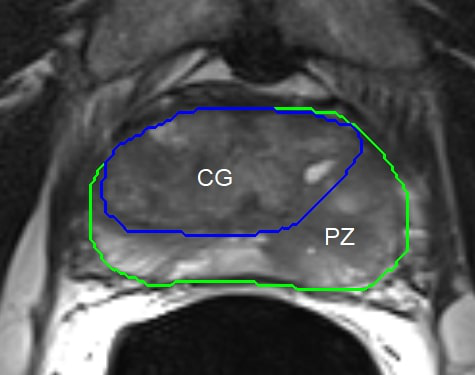}
    \hspace{30pt}
   \includegraphics[width=160pt,height=10.8pc]{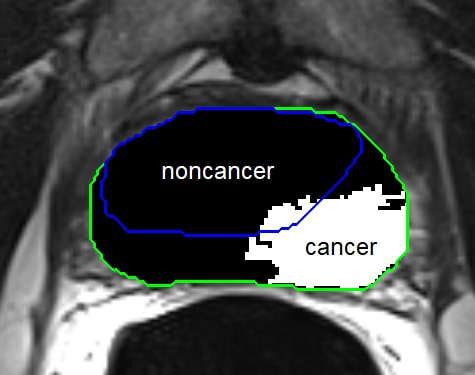}
    } 
\end{center}
	\caption{An example of the manually guided segmentation of the T2-weighted anatomic prostate image in the mpMRI data set. The prostate gland is the area within the green curve. 
	The blue curve demarcates the division between the peripheral zone (PZ) and central gland (CG). Histopathologically identified cancer and noncancer regions are indicated as the white and black areas, respectively, in the second sub-figure.
}\label{fig1}
\end{figure}

Our methods development was motivated by a unique mpMRI data set obtained from patients diagnosed with prostate cancer at the University of Minnesota, the collection procedure for which was previously described (Metzger et al., 2016). 
Briefly, the data were collected on a clinical 3T scanner. The diffusion-weighted and contrast-enhanced images (DWI and DCE-MRI, respectively) were acquired in accordance with Prostate Imaging – Reporting and Data System (PI-RADS) v2 guidelines (Weinreb et al., 2016). Maps of the quantitative MRI parameters were then calculated from these data: apparent diffusion coefficient (ADC) maps were calculated from the DWI data acquired using multiple diffusion-encoding b-values; the maps of DCE-MRI parameters, including the area under the gadolinium concentration time curve at 90 seconds (AUGC90), the forward volume transfer constant ($\text{K}^{\text{trans}}$), and reflux rate constant ($\text{k}_{\text{ep}}$), were generated using a modified Tofts model (Tofts, 1997). Maps of these quantitative parameters were then manually co-registered. Patients that were imaged subsequently underwent radical prostatectomy, and the ex vivo prostate specimens were collected and processed after surgery. The histopathology slides were annotated for cancer by trained pathologists, then co-registered with the quantitative MR maps using a registration method in Kalavagunta et al. (2015). The data contains 34 prostate images, each has $2098\sim5756$ voxels and is from a different patient/subject. 
Figure~\ref{fig1} presents an example of a prostate image showing the two regions of the prostate and areas of cancer and non-cancer.

The general notations for the various data components are as follows. Assume there are $N$ images (subjects), with $n_i$ voxels in the $i^{th}$ image, $i=1,2,\ldots,N$. 
For the $j^{th}$ voxel in the $i^{th}$ image, we have a $d\times 1$ vector of observed mpMRI parameters, $\bm{y_{ij}}=(y_{ij,1},...,y_{ij,d})^T$, which, for our data set, includes the aforementioned ADC, AUGC90, $\text{K}^{\text{trans}}$ and $\text{k}_{\text{ep}}$. 
Note that the ADC values are approximately normally distributed, while the right-skewed AUGC90, $\text{K}^{\text{trans}}$ and $\text{k}_{\text{ep}}$ values are log transformed to have an approximately normal distribution. 
Each voxel has a binary cancer indicator obtained from the co-registered pathology data, 
which we call $c_{ij}$, 
where $c_{ij}=1$ indicates cancer and $0$ indicates noncancer.
In addition, each voxel is identified by a set of location information, including $r_{ij}$, a binary indicator of prostate region, with $r_{ij}= 1$ indicating PZ and $0$ indicating CG, and $\bm{s_{ij}}$, a standardized 2-D coordinate.
Unlike the brain for which a common template already exists, 
the size of the prostate can vary substantially across patients, making it difficult to develop a common template for the prostate. 
Therefore, we linearly rescaled the original 2-D coordinates of the voxels so that they have a common support of $(-1,1)\times(-1,1)$, with $(0,0)$ being the center of each image. 
We denote the set of data components as $\bm{Y}=\{\bm{y_{ij}}| i=1,\ldots,N,j=1,\ldots,n_i\}$, $\bm{c}=\{c_{ij}|i=1,\ldots,N,j=1,\ldots,n_i\}$, $\bm{R}=\{r_{ij}|i=1,\ldots,N,j=1,\ldots,n_i\}$ and  $\bm{S}=\{\bm{s_{ij}}|i=1,\ldots,N,j=1,\ldots,n_i\}$.
Finally, we use $\bm{Y^*}$, $\bm{R^*}$ and $\bm{S^*}$ to denote the set of mpMRI parameters, region information and coordinates, respectively, for voxels in a new prostate image whose cancer status $\bm{c^*}$ are unobserved and must be predicted.

\section{Method}\label{method}
\subsection{Baseline model}\label{baseline}
The classification models proposed in this paper build upon the work of Jin et al. (2018), which focused on modeling the regional heterogeneity in the mpMRI data, i.e. 
the difference in both the distribution of mpMRI parameters and cancer prevalence between the two regions of the prostate, PZ and CG. 
Jin et al. (2018) proposed to model the region-specific joint density of the mpMRI parameters, $\bm{Y}$, and cancer status, $\bm{c}$, given region information, $\bm{R}$, via a Bayesian hierarchical modeling framework. The unknown cancer status for the voxels in a new prostate image, $\bm{c^*}$, are classified using posterior predictive cancer probabilities given $\bm{Y^*}$ and $\bm{R^*}$. 
Specifically, the joint distribution of $\bm{Y}$ and $\bm{c}$ is assumed to depend on $\bm{R}$, which can be defined hierarchically as $f(\bm{Y},\bm{c}|\bm{R}) = f(\bm{Y}|\bm{c},\bm{R})p(\bm{c}|\bm{R})$.  
For $f(\bm{Y}|\bm{c},\bm{R})$: $\bm{y_{ij}}$'s are assumed to follow independent multivariate normal distributions with mean and covariance matrix varying by cancer status $c_{ij}\in\{0,1\}$ and region indicator $r_{ij}\in\{0,1\}$. For $p(\bm{c}|\bm{R})$: $c_{ij}$ is assumed to follow a Bernoulli distribution, $Ber(p_{r_{ij}})$, with cancer probability $p_{r_{ij}}$ varying by region indicator $r_{ij}\in \{0,1\}$:
\begin{align}
\bm{y_{ij}}|c_{ij},r_{ij}
& \stackrel{iid}{\sim}
\mathcal{MVN}(\bm{\mu_{c_{ij},r_{ij}}},\bm{\Gamma_{c_{ij},r_{ij}}}),\nonumber\\ c_{ij}|r_{ij} & \stackrel{iid}{\sim} Ber(p_{r_{ij}}).
\label{model1}
\end{align}
Model (\ref{model1}) is the basis for the development of our proposed models, and will be used as the baseline for evaluating the performance of the models defined below.

\subsection{General modeling approach for the between-patient heterogeneity and spatial correlation}\label{general}
In this sub-section, we introduce our general approach for modeling the between-patient heterogeneity in mpMRI parameters and spatial correlation in the voxel-wise cancer status.

\subsubsection{Between-patient heterogeneity}
Building on the hierarchical structure of the baseline model (\ref{model1}), the between-patient heterogeneity in the mpMRI parameters can be incorporated by introducing a random intercept, 
$\bm{\delta_i}\sim \mathcal{MVN}(\bm{0},\bm{\Sigma})$, on $\{\bm{y_{ij}}|j=1,2,\ldots,n_i\}$, so that $\bm{y_{ij}}|c_{ij},r_{ij},\bm{\delta_i} \stackrel{iid}{\sim} \mathcal{MVN}( \bm{\mu_{c_{ij},r_{ij}}} + \bm{\delta_i}, \bm{\Gamma_{c_{ij},r_{ij}}})$.
In our setting, $\bm{\delta_i}$ represents the subject-specific random shift of the $i^{th}$ patient from the overall mean, with respect to the mpMRI parameters.

\subsubsection{General Gaussian process model for spatial correlation}
The baseline model (\ref{model1}) assumes that all voxels are independent, with voxel-wise cancer probability $p_{r_{ij}}$ only depending on the region indicator of the voxel. 
In reality, however, substantial spatial correlation has been observed in the voxel-wise cancer status within each image. 
To account for this feature, we instead assume that the voxel-wise cancer probabilities in the $i^{th}$ image, which we now denote as $p_{ij}$'s, 
vary by the location/coordinate of the voxels, $\bm{s_{ij}}$'s, 
and are spatially correlated according to the distance between $\bm{s_{ij}}$'s. 
Note that this only implies spatial correlation between voxels within an image and that voxels from different images/patients are assumed independent. 
For Gaussian distributed geostatistical data, Gaussian process models are widely used to model the spatial correlation assuming correlated Gaussian spatial random effects. The $p_{ij}$'s, however, are restricted to the unit interval, and Gaussian spatial random effects cannot be directly applied. Instead, we conduct a probit transformation on $p_{ij}$ and obtain $q_{ij}=\Phi^{-1}(p_{ij})$, where $\Phi^{-1}$ denotes the probit function, 
which has a support of $(-\infty,+\infty)$, 
then 
assume a vector of spatially correlated Gaussian random effects 
$\bm{w_i}=(w_{i1},w_{i2},\ldots,w_{i,n_i})^T$ on $\bm{q_i}=(q_{i1},q_{i2},\ldots,q_{i,n_i})^T$. 

Under the probit model structure, 
we introduce a latent variable, $\kappa_{ij} \sim N(q_{r_{ij},0} + w_{ij} ,1)$, where $q_{r_{ij},0}$ denotes the probit of the cancer prevalence in region $r_{ij}\in \{0,1\}$.
The cancer status, $c_{ij}$, is then defined as $I(\kappa_{ij}>0)$.  
The full model that accounts for both between-patient heterogeneity and spatial correlation between voxels becomes:
\begin{align}
\bm{y_{ij}} \stackrel{iid}{\sim} \mathcal{MVN}(\bm{\mu_{c_{ij},r_{ij}}} + & \bm{\delta_i}, \bm{\Gamma_{c_{ij},r_{ij}}}), \text{\hspace{5pt}}
\bm{\delta_i} \sim \mathcal{MVN}(\bm{0},\bm{\Sigma}),\nonumber \\
c_{ij} = I (\kappa_{ij}>0), \text{\hspace{5pt}}
\kappa_{ij}  {\sim} N(q_{r_{ij},0} & + w_{ij},1),\text{\hspace{5pt}}
\bm{w_i} {\sim} 
\mathcal{MVN}(\bm{0},\bm{C}(\bm{S_i},\bm{S_i}|\bm{\theta})),
\label{generalmodel}
\end{align}
where $\bm{S_i}=(\bm{s_{i1}},\bm{s_{i2}},\ldots,\bm{s_{i,n_i}})^T$, $\bm{C}(\bm{S_i},\bm{S_i}|\bm{\theta})$ denotes the spatial covariance matrix of $\bm{w_i}$, and $\bm{\theta}$ denotes the set of spatial parameters shared by all images.
For the construction of $\bm{C}(\bm{S_i},\bm{S_i}|\bm{\theta})$, 
we assume that $\bm{w_i}$ is the realization of a zero-mean Gaussian process $\text{GP}(\bm{0},\bm{C}(\cdot,\cdot|\bm{\theta}))$ on $\bm{S_i}$.  
We define $\bm{C}(\bm{S_i},\bm{S_i}|\bm{\theta})
=\sigma^2\rho(\bm{S_i},\bm{S_i}|\bm{\zeta})$, where
$\rho(\cdot,\cdot|\bm{\zeta})$ is a correlation function with a set of correlation parameters $\bm{\zeta}$, 
$\sigma^2$ denotes the spatial variance, and $\bm{\theta}=\{\sigma^2,\bm{\zeta}\}$.
In this paper, we employ a Mat\'ern correlation
function (Stein, 2012), which is one of the most widely used correlation functions in spatial statistics that covers various types of stationary spatial correlation patterns. Given the Mat\'ern correlation function, the spatial correlation between two locations $\bm{s},\bm{t}\in \mathbb{R}^2$ is defined as:
\begin{align}
\rho(\bm{s},\bm{t}|\bm{\zeta})=\frac{1}{2^{\nu-1} \Gamma({\nu})} \Big(\frac{2\nu^{1/2}\norm{\bm{s}-\bm{t}}}{\phi}\Big)^{\nu}\bm{J}_{\nu} \Big(\frac{2\nu^{1/2}\norm{\bm{s}-\bm{t}}}{\phi}\Big),
\label{maternfunction}
\end{align}
where $\bm{\zeta}=\{\phi,\nu\}$, $\phi$ is the spatial range parameter with larger $\phi$ indicating larger-scale spatial correlation, 
$\nu$ is the smoothness parameter controlling the degree of differentiability, with larger $\nu$ indicating smoother correlation, $\Gamma(\cdot)$ is the gamma function, $\bm{J}_{\nu}(\cdot)$ is the modified Bessel function of the second kind with order $\nu$, and $\norm{\cdot}$ denotes the Euclidean distance. 

\subsection{Computationally efficient modeling approaches for spatial correlation}\label{efficientspatial}
The full spatial process model (\ref{generalmodel}) becomes 
computationally infeasible on our motivating data set, where there are a large number of images, each involving a separate spatial process over thousands of voxels. In this sub-section, we propose 
three different approaches that ensure scalable spatial modeling of the mpMRI data. 

\subsubsection{Nearest Neighbor Gaussian Process (NNGP)}\label{nngp}
Our first scalable approach to modeling spatial correlation 
is based on sparse approximation via NNGP. 
Most sparse approximation approaches do not necessarily define a valid spatial process, and prediction is through interpolation from a different spatial process that may not reflect the true predictive uncertainty. 
To deal with issue, Datta et al. (2016) proposed NNGP for fully process-based modeling of large spatial data sets, which was shown to significantly outperform competing approaches in terms of inference and scalability. 

The construction of NNGP was discussed in detail by Datta et al. (2016), with applications to spatially-correlated, normally distributed observations on a single map in the linear regression model framework. In our setting, we extend NNGP to Bayesian hierarchical modeling of multi-image data with spatially-correlated binary outcomes. 
We apply a separate NNGP to each image. 
Take the $i^{th}$ image as an example: 
the joint density of $\bm{w_i}$ (i.e. the vector of spatial random effects on $\bm{q_i}$) 
can be written as the product of conditional densities:
\begin{align}
f(\bm{w_i}) = \prod_{j=1}^{n_i} 
f(w_{ij}|w_{i1},w_{i2},\ldots,w_{i,j-1}).
\label{originaldensity}
\end{align}
To reduce the computational burden, 
we replace each large conditioning set $\{w_{i1},w_{i2},\ldots,w_{i,j-1}\}$ with a smaller set of size at most $m$ on $N(\bm{s_{ij}})\subseteq \bm{S_i}\setminus \{ \bm{s_{ij}} \}$, where $m \ll \underset{i}{\text{min }} n_i$, to construct an alternative density: 
\begin{align}
\widetilde{f}(\bm{w_{i}}) = \prod_{j=1}^{n_i} 
f(w_{ij}|\bm{w_{N(\bm{s_{ij}})}}),
\label{alternativedensity}
\end{align}
where $\bm{w_{N(\bm{s_{ij}})}}$ denotes the vector of spatial random effects on  $N(\bm{s_{ij}})$, the $m$ nearest neighbors of $\bm{s_{ij}}$. 
For each image $i$, we view $\{ \bm{S_i}, N_{\bm{S_i}} \}$ as a directed graph $\mathcal{G}$, with $\bm{S_i}$ being the set of nodes and $\bm{N}_{\bm{S_i}}$ the set of directed edges. 
It has been proven that if $\mathcal{G}$ is a directed acyclic graph, then $\widetilde{f}(\bm{w_i})$ in (\ref{alternativedensity}) will be a proper multivariate joint density. 
Specifically, let $C_{\bm{s_{ij}}}$ denote the variance of $w_{ij}$, $\bm{C_{N(\bm{s_{ij}})}}$ the $m \times m$ covariance matrix of $\bm{w_{N(s_{ij})}}$, and $\bm{C_{s_{ij},N(s_{ij})}}$ the $1 \times m$  cross-covariance matrix between $w_{ij}$ and $\bm{w_{N(s_{ij})}}$. We can show that 
\begin{align}
\widetilde{f}(\bm{w_{i}}) = \prod_{j=1}^{n_i} N(w_{ij}| \bm{B_{s_{ij}}} \bm{w_{N(s_{ij})}},F_{\bm{s_{ij}}}), 
\label{alternative2}
\end{align}
where $\bm{B_{s_{ij}}} = \bm{C_{s_{ij},N(s_{ij})} C^{-1}_{N(s_{ij})}}$, $F_{\bm{s_{ij}}} = C_{\bm{s_{ij}}}-\bm{C_{s_{ij},N(s_{ij})} C^{-1}_{N(s_{ij})} C_{N(s_{ij}),s_{ij}}}$. 
In fact, 
$\widetilde{f}(\bm{w_{i}})$ is the probability density function of a multivariate normal distribution, which we denote as $\mathcal{MVN}(\bm{0},\bm{\widetilde{C}_{S_i}})$. 
Given that each $N(\bm{s_{ij}})$ has at most $m$ ($m\ll \underset{i}{\text{min }} n_i$) members, it can be shown that the precision matrix $\bm{\widetilde{C}_{S_i}}^{-1}$ is sparse with at most $ m (m+1)n_i/2$ non-zero entries. 

To construct $\{N(\bm{s_{ij}})|j=1,\ldots,n_i\}$, i.e. the neighbor sets in the $i^{th}$ image, 
we first order the voxels by x-coordinate then y-coordinate, and denote the ordered voxels as $\bm{s_{i1}},\bm{s_{i2}},...,\bm{s_{i,n_i}}$, then define $N(\bm{s_{ij}})$ as the set of $m$ voxels in $\{\bm{s_{i1}},\bm{s_{i2}},...,\bm{s_{i,j-1}}\}$ that have the shortest Euclidean distance from $\bm{s_{ij}}$. 
The ordering of voxels has been shown to have no discernible impact on the approximation of (\ref{originaldensity}) by (\ref{alternativedensity}).
The choice of $m$ can follow standard model comparison metrics such as DIC, but typically a small value between $10 \sim 15$ can provide inference almost indistinguishable to full spatial models for an image with thousands of voxels (Datta et al., 2016). 
Our proposed model with NNGP for spatial modeling still follows the structure of the full model (\ref{generalmodel}), except that we replace $\mathcal{MVN}(\bm{0},\bm{C}_{\bm{S_i}})$, the original prior for $\bm{w_i}$, by the NNGP prior $\mathcal{MVN}(\bm{0},\bm{\widetilde{C}}_{\bm{S_i}})$. 

\subsubsection{Knot-based reduced-rank approximation}
The NNGP-based approach proposed in Section \ref{nngp} reduces computational intensity by inducing sparsity in the large spatial precision matrix, which has been proven to perform well in capturing local spatial dependence structures. 
An alternative approach is through reduced-rank approximation, which is better equipped to capture large-scale, global spatial dependency
(Finley et al., 2009). 

Among the various reduced-rank approximation techniques, we considered a knot-based method proposed by Banerjee et al. (2008), which regresses the original process on its realizations over a smaller set of locations, referred to as ``knots''. 
Take the $i^{th}$ image as an example: we first select a set of $a$ knots, $\bm{S_i^*}=\{\bm{s^*_{i,1}},...,\bm{s^*_{i,a}}\}\subset \bm{S_i}$, where $a \ll \underset{i}{\text{min }} n_i$, with corresponding spatial random effects $\bm{w^*_i}=(w_{\bm{s^*_{i,1}}},...,w_{\bm{s^*_{i,a}}})^T$. 
The original Gaussian process in model (\ref{generalmodel}) yields $\bm{w^*_i}\sim \mathcal{MVN}(\bm{0},\bm{C_{S^*_i}})$, where $\bm{C_{S^*_i}}=\bm{C}(\bm{S^*_{i}},\bm{S^{*}_{i}}|\bm{\theta})$. 
Using $\bm{w^*_i}$ as the basis, for any single site $\bm{s_{ik}}$, the corresponding spatial interpolant is given by $\widetilde{w}(\bm{s_{ik}})=E[w(\bm{s_{ik}})|\bm{w^*_i}]=\bm{C_{s_{ik},S^*_i}}\bm{C_{S_i^*}^{-1}}\bm{w^*_i}$, where $\bm{C_{s_{ik},S^*_i}}=\bm{C}(\bm{s_{ik}},\bm{S^*_{i}}|\bm{\theta})$ is the $1\times a$ cross-covariance matrix between $\bm{w_{s_{ik}}}$ and $\bm{w^*_{i}}$.
This single-site interpolator defines a new spatial process 
$\bm{\text{\tsup{w}}}(\bm{S_i})\sim \text{\textbf{GP}}(\bm{0},\bm{\text{\tsup{C}}}(\cdot,\cdot|\bm{\theta},\bm{S^*_i}))$, where 
$\bm{\text{\tsup{C}}}(\bm{S_i},\bm{S_i}|\bm{\theta},\bm{S^*_i})
=\bm{C_{S_i,S^*_i}}\bm{C_{S_i^*}^{-1}}\bm{C_{S^*_i,S_i}}$, 
and we can replace the original spatial random effects $\bm{w_i}$ by $\bm{\text{\tsup{w}}_i}=E[\bm{w_i}|\bm{w^*_i}]=\bm{C_{S_i,S^*_i}}\bm{C_{S^*_i}^{-1}}\bm{w^*_i}$. 
Since the resulting covariance matrix has a fixed rank $a$ much smaller than $\underset{i}{\text{min }} n_i$, faster computation can be achieved by avoiding inverting spatial covariance matrices of size larger than $a\times a$.

\subsubsection{CAR model}\label{car}
The final approach we consider uses 
Gaussian Markov random fields (GMRF). 
Different from the NNGP-based and reduced-rank approaches, GMRF does not specify a spatial correlation function. Instead, spatial dependency is introduced by specifying the conditional distributions 
$\{f(w_{ij}|\bm{w_{i,-j}})|j=1,\ldots,n_i\}$. In this paper, we apply a popular application of GMRF: the conditional autoregressive (CAR) model, on $\bm{w_i}$:
\begin{align}
w_{ij}|\bm{w_{i,-j}} \sim N(\sum_{k\neq j} b_{ijk} w_{ik}, \sigma^2),
\end{align}
where $b_{ijk}$ is the weight of $w_{ik}$ on $w_{ij}$ that can be specified by the user, and we define $b_{ijk}=\frac{d_{ijk}^{-1}}{\sum_{l \neq j} d_{ijl}^{-1}}$, 
with $d_{ijk}$ denoting the Euclidean distance between voxel $j$ and $k$ in the $i^{th}$ image. Using Brook's lemma (Rue and Held, 2005), we can obtain that $\bm{w_i}\sim \mathcal{MVN}(\bm{0},\sigma^2(\bm{I}-\bm{B_i})^{-1})$, where $\bm{B_i}=[b_{ijk}]_{j,k=1}^{n_i}$. 
The precision matrix of $\bm{w_i}$ has a closed form $\sigma^{-2}\bm{(I-B_i)}$, therefore we have avoided the need to invert large covariance matrices. However, since GMRF does not allow inference on the underlying spatial process, this model may not reveal the true spatial dependence structure, which may limit accuracy of the resulting classifier.

\subsection{MCMC algorithm for Bayesian inference and classification}\label{mcmc}
Bayesian inference and classification of the various models were implemented using MCMC algorithms via Gibbs sampler with a Metropolis-Hastings sampling step. In particular, the non-spatial parameters, as well as $w_{ij}$'s and $k_{ij}$'s, were updated via Gibbs sampling, while the spatial parameters in $\bm{\theta}$ were updated in block via Metropolis-Hastings sampling. 
Our initial simulation results showed that when the spatial variance $\sigma^2$ was large (e.g. 50) and the spatial range $\phi$ was large (e.g. 5), given a flat prior, $q_{r,0}$'s, the probit of the region-specific cancer prevalences, converged to values close to 0 or 1, which were far from their true values. This was possibly due to the large-scale spatial correlation and large spatial variance, which resulted in a small effective sample size for estimating the overall cancer prevalence. Therefore, instead of updating $q_{r,0}$'s along with the other parameters, we set them equal to the sample prevalence $\Phi^{-1}(\sum_{i=1}^{N}\sum_{j:r_{ij}=r}c_{ij}/\sum_{i=1}^{N}\sum_{j:r_{ij}=r}1)$, $r=0,1$. Our simulation results indicate that this substitution does not degrade our ability to appropriately model the other model parameters. 
To facilitate computational efficiency, we conducted a partial parallelization strategy in updating the voxel-level spatial components, including $w_{ij}$'s, $k_{ij}$'s, and the corresponding spatial matrices involved in their full conditional distributions, given the conditional independence between images. Details of the MCMC algorithm are provided in the Web Appendix.

\section{Simulation studies}\label{simulation}
\subsection{Simulation settings}
\begin{figure}[ht!]
\centering
\includegraphics[scale=0.45] {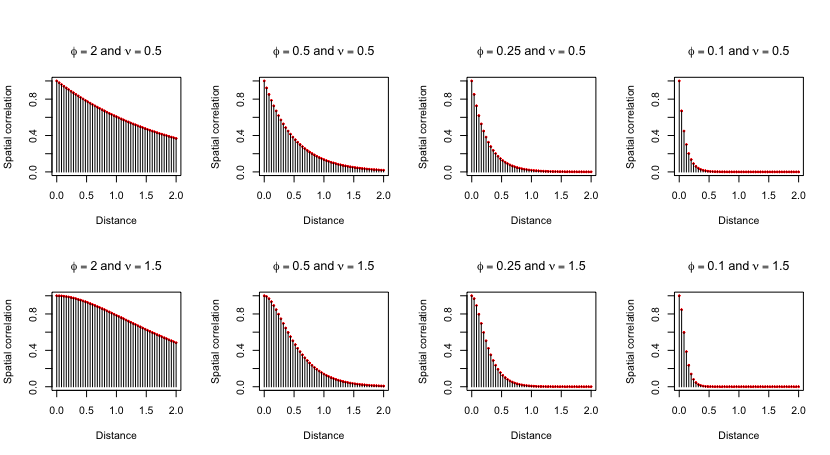}
\caption{Mat\'ern correlation versus spatial distance assuming different values for $\phi$ (range parameter, larger $\phi$ indicates larger-scale correlation) and $\nu$ (smoothness parameter, smaller $\nu$ indicates larger differentiability). The difference in x-axes between two neighboring vertical lines is equal to 0.04, which is the average distance between two neighboring voxels in the motivating data set. The range of the x-axes is $(0,2)$, which is the scale of the pairwise distances between voxels in the motivating data set.}
\label{matern}
\end{figure}

We conducted simulation studies to evaluate the performance of the models discussed in Section \ref{method}. 
To simulate a prostate image, a mask (shape of the prostate image, including the voxel-wise region indicators $r_{ij}$'s and 2-D coordinates $\bm{s_{ij}}$'s) was sampled with replacement from the prostate images in our motivating data set. 
Voxel-wise mpMRI parameters and cancer status were then simulated according to the full model (\ref{generalmodel}), with a spatial correlation structure following the Mat\'ern correlation function (\ref{maternfunction}).
We then applied the proposed models, and compared their performance with the baseline model (\ref{model1}) to evaluate the improvement in classification due to modeling the between-patient heterogeneity and spatial correlation. 
We also compared our proposed models 
with the full model (\ref{generalmodel}), to evaluate their performance in approximating the classification accuracy of full spatial modeling.
Since the full model is computationally infeasible on the original-size images, we generated reduced-size images by taking every third row and 
column of the full-size images, resulting in about $300 \sim 800$ voxels per image. 
The general model parameters, including the mean, within-patient and between-patient covariances of the mpMRI parameters, and the region-specific cancer prevalences, were set equal to the estimates from the motivating data set.
We set $\sigma^2$ (spatial variance) to be 1, 5 or 20, $\phi$ (range parameter
) to be 0.1, 0.25, 0.5 or 2, and $\nu$ (smoothness parameter
) to be 0.5 or 1.5. Figure~\ref{matern} shows the spatial correlation pattern 
given different values of $\phi$ and $\nu$. 
In each simulation, we trained the classifiers using 34 training images, and evaluated their performance on 10 test images. 

\subsection{Simulation results}\label{simuresults}
Table \ref{table1} reports the performance 
of seven candidate models, including M-base: the baseline model; M-sse: M-base plus subject specific effects (SSE) accounting for between-patient heterogeneity in the mpMRI parameters; M-sse-nngp, M-sse-rr, M-sse-car and M-sse-full: M-sse plus spatial modeling using NNGP, reduced-rank approximation, CAR model, and full model, respectively; and M-smooth: the ``Msmooth'' model proposed by Jin et al. (2018), which applies a spatial Gaussian kernel smoother to the posterior predictive cancer probabilities of M-base to account for spatial correlation between voxels, but without accounting for between-patient variability. 
We set $m$ (the number of nearest neighbors in NNGP) and $a$ (the number of knots in reduced-rank approximation) both equal to 10 to ensure a fair comparison between the NNGP and reduced-rank models. The nearest neighbor sets for NNGP were constructed following the procedure introduced in Section \ref{nngp}, and for the reduced-rank model, the knots were selected as a set of equally spaced grids in each image.
All samplers were programmed using the R package ``Rcpp'' (Eddelbuettel et al., 2011). 
\begin{table}[htbp!]
\begin{center}
\caption{Performance of the candidate models given different values for $\sigma^2$ (spatial variance), $\phi$ (spatial range parameter), and $\nu$ (spatial smoothness parameter). 
``M-base'': the baseline model (\ref{model1}) in Section \ref{baseline}; ``M-sse'': the baseline model plus subject-specific effects (SSE) accounting for between-patient heterogeneity in the mpMRI parameters; ``M-sse-nngp'', ``M-sse-rr'', ``M-sse-car'' and ``M-sse-full'': models that account for patient heterogeneity, as well as spatial correlation using NNGP, reduced-rank model, CAR model and full spatial model, respectively.
``M-smooth'': the ``Msmooth'' model in Jin et al. (2018), which is M-base plus a post hoc spatial Gaussian kernel smoother applied to the posterior predictive cancer probabilities. 
Bayesian inference and classification were based on two chains of 25000 MCMC iterations after a burn-in stage of 5000 iterations.
AUCs are summarized as means with standard deviations in the parentheses, which were obtained from 100 simulations for each data scenario.
The ``Time'' row lists the average computation time in hours for each simulation.
\label{table1}
}
\begin{threeparttable} 	
    \footnotesize{}
    \begin{tabular}{cccccccccc}
			\hline\hline
			\multicolumn{3} {c} {Parameters} & \multicolumn{7} {c}{AUC} \\
            \hline
    $\sigma^2$ & $\phi$  &$\nu$ & M-base & M-sse & M-sse-nngp & M-sse-rr & M-sse-car & M-sse-full & M-smooth\\ 
			\hline 
			\multirow{8}{*}{1} & \multirow{2}{*}{2.0} & 0.5 & .755 (.023) & .820 (.015)&.847 (.011) &.851 (.012) &.835 (.010) & .852 (.011) & .562 (.058)\\ %
			& & 1.5 & .750 (.032)  & .823 (.020) & .833 (.012) & .851 (.014) & .837 (.013) & .851 (.013)& .520 (.064) \\ 
			& \multirow{2}{*}{0.5} & 0.5 & .766 (.022) & .819 (.017) & .859 (.011) & .852 (.011) & .834 (.010) & .861 (.011) & .611 (.046)\\ 
			& & 1.5 & .763 (.021) & .819 (.019) & .849 (.011) & .872 (.011) & .838 (.011) & .876 (.010) & .636 (.048)\\ 
            & \multirow{2}{*}{0.25} & 0.5 & .770 (.015) & .815 (.018) & .849 (.010) & .800 (.019) & .830 (.010) & .850 (.011) & .598 (.036) \\ %
			& & 1.5 & .767 (.019) & .820 (.011) & .858 (.010) & .822 (.016) & .833 (.010) & .865 (.010) & .622 (.039) \\ %
            & \multirow{2}{*}{0.1} & 0.5 & .769 (.016) & .817 (.013) & .834 (.007) & .627 (.014) & .827 (.008) & .818 (.008) & .553 (.032) \\ %
			& & 1.5 & .768 (.015) & .815 (.017) & .837 (.008) & .600 (.015) & .828 (.008) & .831 (.009) & .564 (.035) \\ %
			\cline{1-10}
			\multirow{8}{*}{5} & \multirow{2}{*}{2.0} & 0.5 & .760 (.013) & .819 (.014)&.888 (.011) &.888 (.013)&.842 (.009)&.892 (.012) & .675 (.052)\\ %
			& & 1.5 & .751 (.027)  & .809 (.027) & .821 (.023) & .883 (024) & .832 (.020) & .882 (.024) & .639 (.067)\\ %
			& \multirow{2}{*}{0.5} & 0.5 & .770 (.017) & .807 (.018) & .892 (.010) & .885 (.012) & .835 (.009) & .896 (.010) & .722 (.036)\\ 
			& & 1.5 & .767 (.025) & .808 (.020) & .906 (.012) & .919 (.011) & .841 (.010) & .925 (.011) & .778 (.039)\\ 
            & \multirow{2}{*}{0.25} & 0.5 & .776 (.015) & .812 (.011) & .876 (.005) & .848 (.010) & .835 (.005) & .880 (.005) &.693 (.024) \\ %
			& & 1.5 & .777 (.013) & .808 (.021) & .901 (.008) & .873 (.009) & .835 (.008) & .907 (.008) & .755 (.028) \\ %
            & \multirow{2}{*}{0.1} & 0.5 & .777 (.016) & .803 (.017) & .837 (.007) & .688 (.017) & .824 (.006) & .839 (.008) & .607 (.019) \\ %
			& & 1.5 & .778 (.011) & .810 (.009) & .851 (.007) & .688 (.015) & .827 (.007) & .853 (.007) & .644 (.018) \\ %
            \cline{1-10}
			\multirow{8}{*}{20} & \multirow{2}{*}{2.0} & 0.5 & .765 (.012) & .811 (.010)&.902 (.009) &.906 (.010) &.839 (.009) &.910 (.009) & .750 (.054)\\ %
			& & 1.5 & .755 (.027)  & .811 (.024) & .843 (.022) & .923 (025) & .843 (.018) & .924 (.026) & .754 (.076) \\ %
			& \multirow{2}{*}{0.5} & 0.5 & .771 (.015) & .802 (.014) & .903 (.010) & .897 (.013) & .836 (.008) & .914 (.010) & .770 (.032)\\ %
			& & 1.5 & .770 (.021) & .805 (.020) & .940 (.010) & .945 (.011) & .844 (.010) & .952 (.010) & .843 (.034)\\ 
            & \multirow{2}{*}{0.25} & 0.5 & .782 (.009) & .801 (.016) & .882 (.009) & .856 (.013) & .833 (.006) & .893 (.009) & .733 (.025) \\ %
			& & 1.5 & .761 (.026) & .830 (.014) & .924 (.012) & .918 (.015) & .849 (.012) & .925 (.014) & .790 (.027) \\
            & \multirow{2}{*}{0.1} & 0.5 & .782 (.013) & .804 (.014) & .837 (.009) & .698 (.021) & .824 (.008) & .845 (.010) & .629 (.015) \\ %
			& & 1.5 & .786 (.009) & .808 (.011) & .852 (.008) & .710 (.018) & .826 (.007) & .860 (.008) & .673 (.017) \\
			\hline
			\multicolumn{3} {c} {Time} & 0.31 & 0.35 & 2.39 & 3.75 & 0.77 & \rule{0.3cm}{0.5pt} & 0.34 \\
			\hline
		\end{tabular}
        \begin{tablenotes}
        \item[] Note: M-sse-full is computationally intensive for simulation studies if conducted under a fully Bayesian framework. Instead, we fixed the spatial parameters to their true values, and therefore the corresponding computation time is not listed here for comparison. 
        \end{tablenotes}
        \end{threeparttable}
\end{center}
\end{table}

From Table \ref{table1}, we observed that the area under the ROC curve (AUC) was substantially improved by formally modeling the patient heterogeneity in the mpMRI parameters (M-sse versus M-base or M-smooth, which uses post-hoc spatial smoothing) in all scenarios. 
We observed substantial improvement in AUC due to modeling the spatial correlation between voxels (the four models with spatial modeling versus M-sse) given large $\sigma^2$ (large spatial variance), large $\phi$ (large-scale correlation) and large $\nu$ (little differentiability). 
Relative performance of the spatial models varied by scenario (see columns 6-9), with $\sigma^2$ having little effect but $\phi$ and $\nu$ having a noticeable effect on the performance. 
The M-sse-nngp model 
had an average AUC close to that of the full model (M-sse-full) when $\phi$ equaled $0.1$ and $0.25$ (small-scale, local correlation) regardless of $\nu$, and also performed well when $\phi$ equaled $0.5$ and $2$ with the exception of the scenario where $\nu = 1.5$. 
The M-sse-rr model, on the other hand, had an average AUC close to that of M-sse-full when $\phi$ equaled $0.5$ or $2$ (larger-scale correlation), 
but performed poorly when $\phi$ equaled $0.1$ or $0.25$ (local correlation), having worse performance than M-base in some scenarios
(e.g. $\sigma^2=1$, $\phi=0.1$ and $\nu=1.5$).
Both M-sse-nngp and M-sse-rr had an average AUC closer to that of M-sse-full when $\nu$ was smaller (larger differentiability).
In general, M-sse-nngp approximated the AUC of M-sse-full well under most data scenarios except when $\phi=2$ and $\nu=1.5$, where there was smooth and very strong correlation across the whole image (see Figure \ref{matern}); in contrast, M-sse-nngp did not approximate the AUC of M-sse-full well unless the spatial correlation had a large scale (e.g. $\phi=2$).
One possible explanation for M-sse-rr having poor performance in some scenarios is that 10 knots is not adequate for the reduced-rank approximation to capture the underlying spatial structure unless there is strong, large-scale spatial correlation across the whole image. In fact, our simulation results indicated that a much larger number of knots ($a\geq 100$) was required for good performance, which would lead to a much longer computation time and thus was not conducted as part of our simulation study.
The M-sse-nngp model, on the other hand, demonstrated its advantage in that a much smaller number of nearest neighbors, in other words, much less computation time, is required, to approximate the classification accuracy of the full model. 
The M-sse-car model had improved AUC compared with M-sse, but overall did not approximate the full model well.
The performance of M-sse-car was not affected by $\sigma^2$, $\phi$ or $\nu$ since the CAR model assumes a fixed spatial structure with no unknown spatial correlation parameters. In summary,
the M-sse-nngp model demonstrated robust performance, offering higher classification accuracy and better approximation to the full model than M-sse-rr and M-sse-car in the majority of the simulated data scenarios.

We also investigated the performance of the classifiers under a more complex spatial pattern, where the spatial covariance follows the average of three different Mat\'ern covariance functions with $(\sigma^2, \phi, \nu)$ equal to $(20,0.25,0.5)$, $(20,1,1)$ and $(20,4,1.5)$, respectively. 
Following the same simulation procedure, the mean and standard deviation (in the parentheses) of AUC were 0.765 (0.021) for M-base, 0.803 (0.022) for M-sse, 0.892 (0.017) for M-sse-nngp, 0.880 (0.018) for M-sse-rr, 0.834 (0.013) for M-sse-car, 0.902 (0.016) for M-sse-full, and 0.769 (0.038) for M-smooth. This relative performance between models is similar to that when 
generating data from a Mat\'ern correlation structure, thus illustrating the robustness of our results to model mis-specification.

\section{Application to in vivo data}\label{application}
We next illustrate the performance of the candidate models on the motivating mpMRI data set described in Section \ref{data}. 
Our preliminary analyses suggest that the between-patient variation in mpMRI parameters has a complex pattern that might be hard to estimate
with a limited sample (34 patients/subjects). 
Therefore, we applied each proposed spatial modeling approach both with and without 
subject specific effects (SSE) accounting for the between-patient variability, which leads to  
M-nnngp and M-sse-nngp: spatial modeling using NNGP without and with SSE; M-rr and M-sse-rr: spatial modeling using reduce-rank approximation without and with SSE; M-car and M-sse-car: spatial modeling using the CAR model without and with SSE.
Summaries of the ROC curve were obtained using 5-fold Cross-Validation to account for over-fitting due to training and evaluating the model on the same data set (Friedman, Hastie and Tibshirani, 2001).

\begin{sidewaystable}[htbp!]
\begin{center}
\caption{Model performance on the motivating mpMRI data set. The candidate models include: ``M-base'': the baseline model (\ref{model1}) in Section \ref{baseline}, ``M-nnngp'' and ``M-sse-nngp'': spatial modeling using NNGP without and with SSE, respectively, ``M-rr'' and ``M-sse-rr'': spatial modeling using reduce-rank approximation without and with SSE, respectively, 
and ``M-car'' and ``M-sse-car'': spatial modeling using the CAR model without and with SSE, respectively. Bayesian inference and classification were based on two chains of 75000 MCMC iterations, after a burn-in stage of 5000 iterations. The ``AUC'' row lists the average AUC obtained from 5-fold Cross-Validation, the ``S80'' row lists the sensitivity corresponding to 80\% specificity, and the ``Time'' row lists the computation time in hours to complete 5-fold Cross-Validation. 
\label{realtable}
}
    \begin{threeparttable} 	
    \begin{tabular}{cccccccc}
			\hline\hline
			&\multirow{2}{*}{M-base} & \multicolumn{2} {c}{NNGP} & \multicolumn{2} {c}{Reduced-rank} & \multicolumn{2} {c}{CAR}\\
     &   & M-nngp & M-sse-nngp & M-rr & M-sse-rr & M-car & M-sse-car \\ 
			\hline 
			AUC & 0.763 & 0.808 & 0.785 & 0.807 & 0.774 & 0.764 & 0.765 \\
            S80 & 0.615 & 0.673 & 0.641 & 0.680 & 0.637 & 0.607 & 0.605 \\	
            $\phi$ & \rule{0.3cm}{0.4pt} &  1.67 (1.49, 1.82)  & 1.64 (1.43, 1.85) & 0.71 (0.60, 0.81) & 0.76 (0.63, 0.85) & \rule{0.3cm}{0.4pt} &\rule{0.3cm}{0.4pt} \\
            $\nu$ & \rule{0.3cm}{0.4pt} & 0.88 (0.85, 0.92) & 0.89 (0.82, 0.98) & 15.89 (13.92, 18.21) & 16.03 (13.97, 18.56) & \rule{0.3cm}{0.4pt} & \rule{0.3cm}{0.4pt} \\
            $\sigma^2$ & \rule{0.3cm}{0.4pt} & 657.9 (619.4, 708.9) & 632.2 (598.3, 694.0) & 765.6 (702.2, 826.1) & 720.5 (562.9, 791.3) 
            & 4.7 (3.5, 6.3) & 4.5 (3.4, 6.3) \\
			Time & 8.52 & 59.78 & 59.96 & 96.91 & 97.19 & 19.02 & 19.24 \\
			\hline
		\end{tabular}
        \begin{tablenotes}
        \item[] Note: 
        Results for the NNGP and reduced-rank models were obtained using 10 nearest neighbors and 10 knots, respectively. The NNGP using more than 10 neighbors gave similar AUC and S80. The reduced-rank model could potentially have improved classification if we use much more knots ($a=100$, for example). However, the MCMC algorithm becomes much more computationally intensive, and in fact is unrealistic to apply to the data.
        \end{tablenotes}
        \end{threeparttable}
\end{center}
\end{sidewaystable}

Table~\ref{realtable} summarizes the performance of the candidate models using AUC, S80 (sensitivity corresponding to 80\% specificity), posterior mean of the spatial parameters and computation time in hours to complete the 5-fold Cross-Validation. Spatial modeling with NNGP and reduced-rank approximation demonstrated improvement in the AUC and S80 both with and without SSE compared with M-base: the AUC increased from 0.763 to 0.808 (M-nngp), 0.785 (M-sse-nngp), 0.807 (M-rr) and 0.774 (M-sse-rr), and the S80 increased from 0.615 to 0.673 (M-nngp), 0.641 (M-sse-nngp), 0.680 (M-rr) and 0.637 (M-sse-rr). This suggests that spatial modeling improved the classification accuracy by successfully modeling the spatial correlation in the data.
The M-car and M-sse-car models, however, had performance essentially equivalent to M-base both without SSE (AUC: 0.764, S80: 0.607) and with SSE (AUC: 0.765, S80: 0.605). 
Adding SSE to account for between-patient heterogeneity on top of spatial modeling did not result in additional improvement in AUC or S80, and, in fact, lowered the AUC and S80 of M-nngp and M-rr. One possible explanation is that the SSE might have a complex distribution, with features that cannot be reflected by the assumed multivariate normal distribution. 

The three proposed spatial modeling approaches lead to different posterior estimates for the spatial parameters. 
For the NNGP-based models, the posterior mean of the spatial range ($\widehat{\phi}$), smoothness ($\widehat{\nu}$) and variance ($\widehat{\sigma^2}$) were 1.67, 0.88 and 657.9, respectively, without SSE, and 1.64, 0.89 and 632.2, respectively, with SSE.
For the models using reduced-rank approximation, $\widehat{\phi}$, $\widehat{\nu}$ and $\widehat{\sigma^2}$ were 0.71, 15.89 and 765.6, respectively, without SSE, and 0.76, 16.03 and 720.5, respectively, with SSE.
For the CAR-based models, there is no assumed $\phi$ or $\nu$, and the posterior mean of $\sigma^2$ was 4.7 without SSE and 4.5 with SSE. 
We can observe that, for each spatial modeling approach, modeling with or without SSE resulted in similar estimated spatial parameters. 
For the models without SSE, M-nngp and M-rr both gave large $\widehat{\sigma^2}$, indicating strong spatial variation among voxels within each image. Although $\widehat{\phi}$ differed (1.67 for M-nngp and $0.71$ for M-rr), both indicated large-scale spatial dependency (see Figure \ref{matern}). The M-nngp model gave a smaller $\widehat{\nu}$ of 0.88, possibly because the NNGP tends to capture local spatial dependencies that are less smooth. The M-rr model, on the other hand, gave a larger $\widehat{\nu}$ of 15.89, indicating a much smoother correlation structure, possibly because reduced-rank approximation tends to capture large-scale, smooth spatial dependencies.
In contrast, the CAR model assumed a fixed spatial dependence structure that was determined by the distance between voxels, which was quite different from the estimated spatial structures using NNGP and reduced-rank approximation, and thus M-car gave a quite different estimated spatial variance $\widehat{\sigma^2}=4.7$. 
There was also substantial variability in the computational intensity. For the 5-fold Cross-Validation with 75000 MCMC iterations after a burn-in stage of 5000 iterations, the NNGP-based models were approximately 3.1 times slower than the CAR-based models, and the models using reduced-rank approximation were approximately 4.9 times slower than the CAR-based models. Computation time for the various models are reported in Table \ref{realtable}.

\begin{sidewaysfigure}
\centering
\includegraphics[scale=0.43] {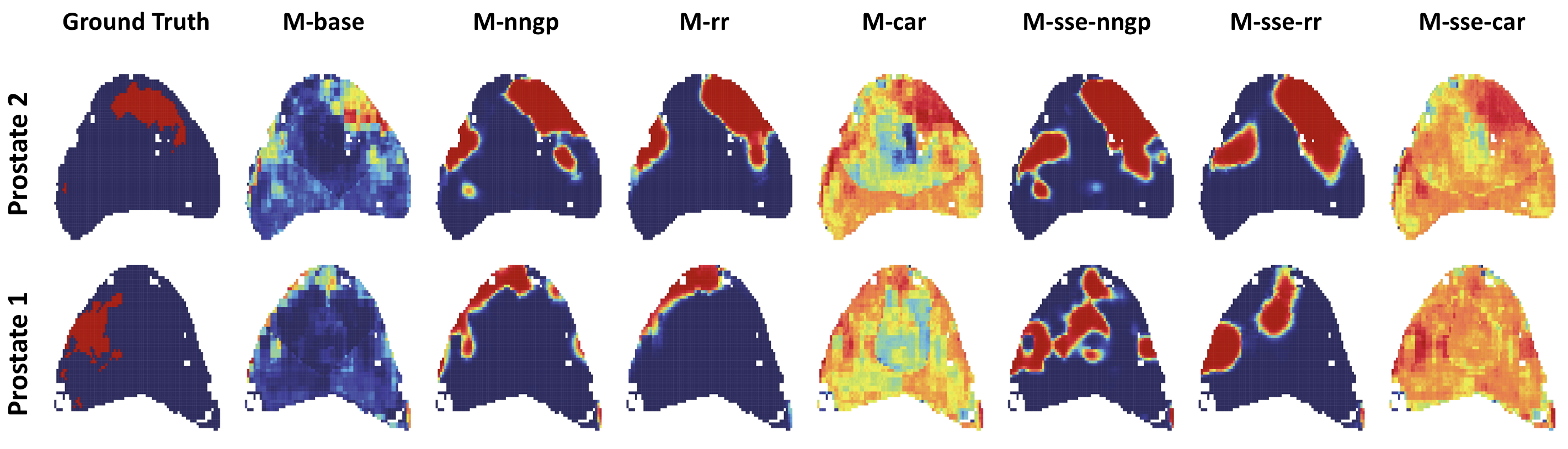} 
\caption{Maps of two representative prostates: ground truth (column 1, red regions represent the registered areas of cancer), heatmaps of posterior predictive cancer probabilities obtained from M-base (baseline model), M-nngp (NNGP without SSE), M-rr (reduced-rank approximation without SSE), M-car (CAR without SSE), M-sse-nngp (NNGP with SSE), M-sse-rr (reduced-rank approximation with SSE), M-sse-car (CAR with SSE), respectively (columns 2-8, where warmer color indicates higher posterior predictive cancer probability, and the color was scaled by the range of the posterior predictive cancer probabilities within each image). The white dots in the heatmaps indicate missing values for at least one mpMRI parameter for the voxels. 
}
\label{heatmap}
\end{sidewaysfigure}

\begin{sidewaysfigure}
\centering
\includegraphics[scale=0.43] {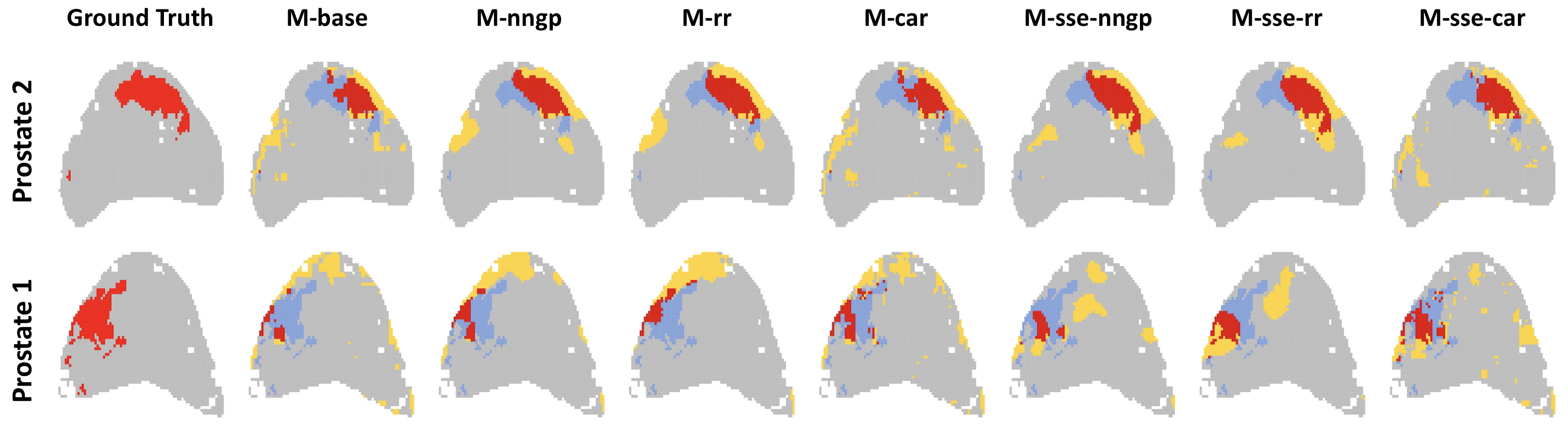} 
\caption{Classification maps of two representative prostates that categorize voxels into true positive (red), false positive (yellow), true negative (grey) and false negative (blue), using the probability cut-off corresponding to 80\% specificity. Models include M-base (baseline model), M-nngp (NNGP without SSE), M-rr (reduced-rank approximation without SSE), M-car (CAR without SSE), M-sse-nngp (NNGP with SSE), M-sse-rr (reduced-rank approximation with SSE), M-sse-car (CAR with SSE), respectively (columns 2-8).}
\label{classificationmap}
\end{sidewaysfigure}

Figures~\ref{heatmap} and~\ref{classificationmap} show heatmaps and classification maps, respectively, for two representative prostate images.
In the heatmaps, warmer color indicates higher posterior predictive cancer probability, and in the classification maps, 
red, yellow, grey and blue indicate correctly identified cancer regions (true positive), incorrectly identified cancer regions (false positive), correctly identified noncancer regions (true negative), and incorrectly identified noncancer regions (false negative), respectively, using the probability cut-off corresponding to 80\% specificity. 
Compared with M-base which generated diffuse areas of high posterior cancer probabilities scattered throughout the images, all spatial models identified larger and more integrated true positive areas. The models using NNGP and reduced-rank approximation produced prediction maps with clearly distinguished cancer and noncancer regions; this is likely because the estimated $\phi$ and $\nu$ for the two models indicate strong, large-scale spatial correlation in the voxel-wise cancer probabilities, which leads to strongly clustered regions of cancer or noncancer voxels. 
Prediction maps from the CAR-based models are much more smooth, and it is hard to distinguish the cancer and noncancer regions by eye. Using a probability cut-off that corresponds to 80\% specificity, however, we can see from Figure~\ref{classificationmap} that the CAR-based models had classification results similar to those of M-base.
In addition, the models that did not account for patient heterogeneity (no SSE) generated prediction maps closer to the ground truth than models with SSE: compared with M-nngp and M-rr, M-sse-nngp and M-sse-rr identified more areas of high posterior cancer probabilities which lead to more true positive areas, but also larger areas of false positives around true positives. 
We have also applied several widely used machine-learning techniques for developing MR classifiers for prostate cancer to our in vivo data, including the random forest (RF), K-nearest neighbors algorithm (KNN), Naive Bayes, and quadratic discriminant analysis (QDA). The corresponding average AUCs obtained from 5-fold Cross-Validation were 0.711, 0.711, 0.724 and 0.729, respectively, which were lower than those presented in Table \ref{realtable}, 
indicating the superiority of our proposed method. 
The advantage of our proposed modeling approach is to be expected: these machine-learning approaches offer tremendous flexibility but likely lead to over-fitting for the small sample sizes typically found in mpMRI studies (Lemaître et al., 2015). 
In contrast, our approach provides the structure that is needed to develop an accurate classifier with small sample sizes, while building in the flexibility needed to target specific features of the data that could improve classification.

\section{Discussion}\label{discussion}
This paper proposes Bayesian hierarchical models for high-resolution mpMRI data, which aim to improve the voxel-wise classification of prostate cancer by modeling the spatial correlation in the voxel-wise cancer status as well as the between-patient variability in the mpMRI parameters. 
To our best knowledge, our method is the first to investigate the modeling of spatial correlation in high-dimensional MR classification.
This enhanced voxel-wise cancer classification can be utilized as a first step of cancer lesion identification for improving prostate cancer management. Initially, the method would greatly impact the clinic through a reader-independent method for targeting biopsies from mpMRI data, with the potential to increase the confidence with which clinically significant cancer is diagnosed thus impacting treatment decisions. In the future, such method may play a role in improved guidance for focal therapies.

Conventional spatial modeling becomes computationally infeasible for voxel-level cancer classification due to the large size of the multi-image mpMRI data. We consider three scalable 
approaches using sparse approximation via NNGP, reduced-rank approximation through a knot-based approach and GMRF with a CAR model. Simulation results indicate that classification can be substantially improved by modeling both the patient heterogeneity in the mpMRI parameters and spatial correlation structure within an image. The proposed M-sse-nngp model performed better under more local, smaller-scale spatial dependency, 
but was robust to the true spatial structure and had the best overall performance. The proposed M-sse-rr model outperformed M-sse-nngp only when there was strong correlation across the whole image. Otherwise, it performed poorly and even worse than non-spatial models when there was only small-scale correlation. 
The proposed M-sse-car model had higher AUC than the non-spatial models, but performed worse than M-sse-nngp in all simulated data scenarios. 
Application to in vivo data showed that spatial modeling using NNGP and reduced-rank approximation improved the average AUC and S80 of M-base, while spatial modeling using the CAR model did not. Modeling between-patient heterogeneity on top of spatial modeling did not further improve the average AUC and S80, and, in some cases, actually decreased performance. This is possibly because the between-patient variation has a complex structure that is hard to estimate only using information from the 34 patients. 
In fact, our preliminary analysis suggests that the distribution of the subject specific effects may be bimodal, and there might be outliers. 
We did investigate more complex modeling strategies, 
which, however, did not show better performance, suggesting that our current data might be insufficient for estimating a more complex random effects distribution. However, we do note that between-patient variation is an important feature of mpMRI, and further investigation should be conducted once more data are available.

The major contribution of the paper is to investigate scalable spatial modeling strategies, which address the computational challenge brought by simultaneously modeling and classifying multiple high-dimensional images with spatial correlation structures. Among the proposed spatial modeling approaches, CAR demonstrated the highest computational efficiency but low classification accuracy.  
This is likely due to the fixed spatial correlation structure used in the CAR model, which might be far from the truth.
Although the NNGP-based and reduced-rank classifiers demonstrated similar classification accuracy on in vivo data, the NNGP-based classifier was approximately 1.6 times faster 
given equal numbers of nearest neighbors and knots ($m=a=10$). 
The computation time increases linearly with the increase in $m$ for NNGP, and increases slightly faster with the increase in $a$ for reduced-rank approximation. Application results showed that the NNGP-based classifier using 10 neighbors gave AUC and S80 almost as high as using more than 10. 
The reduced-rank classifier, however, could potentially have improved accuracy if using many more knots (e.g. $a=100$), but the MCMC algorithm will be too computationally intensive for our data. 
As a general conclusion, the NNGP-based classifier is recommended for application, considering its robust performance with respect to the spatial correlation pattern, high classification accuracy, and the small number of nearest neighbors required that ensures scalability for high-dimensional spatial modeling of the between-voxel correlation. 
The NNGP-based classifiers completed 5-fold Cross Validation with 80000 MCMC iterations per fold in approximately 60 hours on in vivo data. The computation time increases linearly as the number of voxels per image increases, and therefore could be a computational burden in practice. However, this issue could be alleviated by applying alternative approximation algorithms, such as HMC-NUTS (Hoffman and Gelman, 2014), for the posterior distribution.

We currently conduct spatial modeling assuming a stationary correlation structure, while in reality the correlation structure can be more complex. A future extension is to conduct non-stationary spatial modeling. For example, the spatial correlation may change as a function of the location in the prostate, and properly modeling this non-stationary structure could improve performance.
Another future direction is cancer lesion identification using voxel-wise mpMRI data, which has so far received limited attention in the literature (Litjens et al., 2014, Leng et al., 2018).
We are currently discussing voxel-wise classification of prostate cancer, which is the focus for the majority of the existing quantitative mpMRI classifiers. 
However, clinical practice requires that the results are eventually translated into detection of cancer lesions which could better guide the decisions for clinical treatment, and determining how best to identify lesions using voxel-wise data is worthy of investigation.

Our proposed method was motivated by the specific features of prostate mpMRI. 
However, similar challenges will arise in settings that consider other organs or imaging modalities, as well, which also include high-dimensional spatial modeling. In this case, our specific classifiers cannot be directly applied, but our general modeling approach and evaluation of the various approaches to scalable modeling are relevant. 
Among the three proposed scalable spatial modeling approaches, we recommend the NNGP-based approach considering its robustness with respect to spatial correlation pattern, high classification accuracy and scalability. 
Note that the mpMRI parameters should be transformed to ensure the validity of normality assumption, or assumed to have different distributions, which, however, might complicate the modeling procedure. 
Finally, our method was developed for 2-D mpMRI due to the technical challenges related to co-registering the MRI parameters and histopathology in 3-D. However, its extension to 3-D can be easily accomplished by replacing the 2-D coordinates with their 3-D analogs, where the computational complexity only increases linearly with the increase in the number of voxels per image.

\section*{Acknowledgements}
This work was supported by NCIR01 CA155268, NCIP30 CA077598, NIBIBP41 EB015894, and the Assistant Secretary
of Defense for Health affairs, through the Prostate Cancer Research Program under Award No. W81XWH-15-1-0478.
Opinions, interpretations, conclusions, and recommendations are those of the author and are not necessarily endorsed
by the Department of Defense.

\section*{Supporting Information}
The R code for implementing our proposed models is available at\\ \href{https://github.com/Jin93/PCa-mpMRI2}{https://github.com/Jin93/PCa-mpMRI2}.
Additional supporting information may be found online in the Supporting Information Section at the end of the article.




\begin{thebibliography}{}



\bibitem[\protect\citeauthoryear{Not imporant}{2506}]{Banerjee2008}
Banerjee, S., Gelfand, A.E., Finley, A.O. and Sang, H. (2008). Gaussian predictive process models for large spatial data sets.
{\em Journal of the Royal Statistical Society: Series B (Statistical Methodology)} {\bf 70(4)}, 825-848.




\bibitem[\protect\citeauthoryear{Not imporant}{2506}]{Cressie2008}
Cressie, N. and Johannesson, G. (2008). Fixed rank kriging for very large spatial data sets.
{\em Journal of the Royal Statistical Society: Series B (Statistical Methodology)} {\bf 70(1)}, 209-226.

\bibitem[\protect\citeauthoryear{Not imporant}{2506}]{Datta2016}
Datta, A., Banerjee, S., Finley, A.O. and Gelfand, A.E. (2016). Hierarchical nearest-neighbor Gaussian process models for large geostatistical datasets.
{\em Journal of the American Statistical Association} {\bf 111(514)}, 800-812.


\bibitem[\protect\citeauthoryear{Not imporant}{Dickinson2011}]{Dickinson2011}
Dickinson, L., Ahmed, H.U., Allen, C., Barentsz, J.O., Carey, B., Futterer, J.J., et al. (2011). Magnetic resonance imaging for the detection, localisation, and characterization of prostate cancer: recommendations from a European consensus meeting.
{\em European urology} {\bf 59(4)}, 477-494.

\bibitem[\protect\citeauthoryear{Not imporant}{2506}]{Eddelbuettel2011}
Eddelbuettel, D., François, R., Allaire, J., Ushey, K., Kou, Q., Russel, N., et al. (2011). Rcpp: Seamless R and C++ integration.
{\em Journal of Statistical Software} {\bf 40(8)}, 1-18.

\bibitem[\protect\citeauthoryear{Not imporant}{2506}]{Eidsvik2014}
Eidsvik, J., Shaby, B.A., Reich, B.J., Wheeler, M. and Niemi, J. (2014). Estimation and prediction in spatial models with block composite likelihoods.
{\em Journal of Computational and Graphical Statistics} {\bf 23(2)}, 295-315.

\bibitem[\protect\citeauthoryear{Not imporant}{2506}]{Finley2009}
Finley, A.O., Banerjee, S. and McRoberts, R.E. (2009). Hierarchical spatial models for predicting tree species assemblages across large domains.
{\em The annals of applied statistics} {\bf 3(3)}, 1052.

\bibitem[\protect\citeauthoryear{Not imporant}{2506}]{Friedman2001}
Friedman, J., Hastie, T. and Tibshirani, R. (2001). The elements of statistical learning.
{\em New York, NY, USA:: Springer series in statistics} {\bf 1(10)}.






\bibitem[\protect\citeauthoryear{Not imporant}{2506}]{Garcia-Reyes2015}
Garcia-Reyes, K., Passoni, N.M., Palmeri, M.L., Kauffman, C.R., Choudhury, K.R., Polascik, T.J., et al. (2015). Detection of prostate cancer with multiparametric MRI (mpMRI): effect of dedicated reader education on accuracy and confidence of index and anterior cancer diagnosis.
{\em Abdominal imaging} {\bf 40(1)}, 134-142.


\bibitem[\protect\citeauthoryear{Not imporant}{Higdon2002}]{Higdon2002}
Higdon, D. (2002). Space and space-time modeling using process convolutions. 
{\em Quantitative methods for current environmental issues}, 37-56. Springer, London.

\bibitem[\protect\citeauthoryear{Not imporant}{Hoffman2014}]{Hoffman2014}
Hoffman, M. D. and Gelman, A. (2014). The No-U-Turn sampler: adaptively setting path lengths in Hamiltonian Monte Carlo. {\em Journal of Machine Learning Research}, 
{\bf 15(1)}, 1593-1623.




\bibitem[\protect\citeauthoryear{Not imporant}{2506}]{jin2018}
Jin, J., Zhang L., Leng E., Metzger G.J. and Koopmeiners, J.S. (2018).
Detection of prostate cancer with multiparametric MRI utilizing the anatomic structure of the prostate.
{\em Statistics in Medicine} {\bf 46}, 3214--3229.


\bibitem[\protect\citeauthoryear{Not imporant}{2506}]{Kalavagunta2015}
Kalavagunta, C., Zhou, X., Schmechel, S.C. and Metzger, G.J. (2015). Registration of in vivo prostate MRI and pseudo‐whole mount histology using Local Affine Transformations guided by Internal Structures (LATIS).
{\em Journal of Magnetic Resonance Imaging} {\bf 41(4)}, 1104-1114.

\bibitem[\protect\citeauthoryear{Not imporant}{2506}]{kaufman2008}
Kaufman, C.G., Schervish, M.J. and Nychka, D.W. (2008). Covariance tapering for likelihood-based estimation in large spatial data sets.
{\em Journal of the American Statistical Association} {\bf 103(484)}, 1545-1555.

\bibitem[\protect\citeauthoryear{Not imporant}{2506}]{Khalvati2015}
Khalvati, F., Wong, A. and Haider, M.A. (2015). Automated prostate cancer detection via comprehensive multi-parametric magnetic resonance imaging texture feature models. 
{\em BMC medical imaging} {\bf 15(1)}, 27.

\bibitem[\protect\citeauthoryear{Not imporant}{2506}]{Kurhanewicz2008}
Kurhanewicz, J., Vigneron, D., Carroll, P. and Coakley, F. (2008). Multiparametric magnetic resonance imaging in prostate cancer: present and future.
{\em Current opinion in urology} {\bf 18(1)}, 71.


\bibitem[\protect\citeauthoryear{Not imporant}{2506}]{Lemaitre2015}
Lemaître, G., Martí, R., Freixenet, J., Vilanova, J.C., Walker, P.M. and Meriaudeau, F. (2015). Computer-aided detection and diagnosis for prostate cancer based on mono and multi-parametric MRI: a review.
{\em Computers in biology and medicine} {\bf 60}, 8-31.

\bibitem[\protect\citeauthoryear{Not imporant}{2506}]{Leng2018}
Leng, E., Spilseth, B., Zhang, L., Jin, J., Koopmeiners, J.S. and Metzger, G.J. (2018). Development of a measure for evaluating lesion‐wise performance of CAD algorithms in the context of mpMRI detection of prostate cancer.
{\em CMedical physics} {\bf 45(5)}, 2076-2088.

\bibitem[\protect\citeauthoryear{Not imporant}{2506}]{Litjens2014}
Litjens, G., Debats, O., Barentsz, J., Karssemeijer, N. and Huisman, H. (2014). Computer-aided detection of prostate cancer in MRI.
{\em IEEE transactions on medical imaging} {\bf 33(5)}, 1083-1092.

\bibitem[\protect\citeauthoryear{Not imporant}{2506}]{Metzger2016}
Metzger, G.J., Kalavagunta, C., Spilseth, B., Bolan, P.J., Li, X., Hutter, D., et al. (2016). Detection of prostate cancer: quantitative multiparametric MR imaging models developed using registered correlative histopathology.
{\em Radiology} {\bf 279(3)}, 805-816.

\bibitem[\protect\citeauthoryear{Not imporant}{2506}]{Niaf2012}
Niaf, E., Rouvi\`ere, O., M\`ege-Lechevallier, F., Bratan, F. and Lartizien, C. (2012). Computer-aided diagnosis of prostate cancer in the peripheral zone using multiparametric MRI.
{\em Physics in Medicine \& Biology} {\bf 57(12)}, 3833.

\bibitem[\protect\citeauthoryear{Not imporant}{2506}]{Peng2013}
Peng, Y., Jiang, Y., Antic, T., Giger, M.L., Eggener, S. and Oto, A. (2013). A study of T 2-weighted MR image texture features and diffusion-weighted MR image features for computer-aided diagnosis of prostate cancer. In 
{\em Medical Imaging 2013: Computer-Aided Diagnosis} {\bf 8670}, 86701H. International Society for Optics and Photonics.


\bibitem[\protect\citeauthoryear{Not imporant}{2506}]{Rosenkrantz2013}
Rosenkrantz, A.B., Kim, S., Lim, R.P., Hindman, N., Deng, F.M., Babb, J.S., et al. (2013). Prostate cancer localization using multiparametric MR imaging: comparison of Prostate Imaging Reporting and Data System (PI-RADS) and Likert scales.
{\em Radiology} {\bf 269(2)}, 482-492.

\bibitem[\protect\citeauthoryear{Not imporant}{2506}]{Rue2005}
Rue, H. and Held, L. (2005). Gaussian Markov random fields: theory and applications.
{\em CRC press}.



\bibitem[\protect\citeauthoryear{Not imporant}{2506}]{Shah2012}
Shah, V., Turkbey, B., Mani, H., Pang, Y., Pohida, T., Merino, M.J., et al. (2012). Decision support system for localizing prostate cancer based on multiparametric magnetic resonance imaging.
{\em Medical physics} {\bf 39(7Part1)}, 4093-4103.


\bibitem[\protect\citeauthoryear{Not imporant}{Stein2012}]{Stein2012}
Stein, M.L. (2012). Interpolation of spatial data: some theory for kriging.
{\em Springer Science \& Business Media}.


\bibitem[\protect\citeauthoryear{Not imporant}{2506}]{Stein2004}
Stein, M.L., Chi, Z. and Welty, L.J. (2004). Approximating likelihoods for large spatial data sets.
{\em Journal of the Royal Statistical Society: Series B (Statistical Methodology)} {\bf 66(2)}, 275-296.

\bibitem[\protect\citeauthoryear{Not imporant}{2506}]{Tofts1997}
Tofts, P.S. (1997). Modeling tracer kinetics in dynamic Gd‐DTPA MR imaging.
{\em Journal of magnetic resonance imaging} {\bf 7(1)}, 91-101.


\bibitem[\protect\citeauthoryear{Not imporant}{Vecchia1988}]{Vecchia1988}
Vecchia, A.V. (1988). Estimation and model identification for continuous spatial processes.
{\em Journal of the Royal Statistical Society. Series B (Methodological)}, 297-312.


\bibitem[\protect\citeauthoryear{Not imporant}{2506}]{Vos2012}
Vos, P.C., Barentsz, J.O., Karssemeijer, N. and Huisman, H.J. (2012). Automatic computer-aided detection of prostate cancer based on multiparametric magnetic resonance image analysis. 
{\em Physics in Medicine \& Biology} {\bf 57(6)}, 1527.

\bibitem[\protect\citeauthoryear{Not imporant}{2506}]{Weinreb2016}
Weinreb, J.C., Barentsz, J.O., Choyke, P.L., Cornud, F., Haider, M.A., Macura, K.J., et al. (2016). PI-RADS prostate imaging–reporting and data system: 2015, version 2.  
{\em European urology} {\bf 69(1)}, 16-40.



\end{thebibliography}
\end{document}